\documentclass{article}

\usepackage[numbers]{natbib}

\usepackage[preprint]{neurips_2021}

\usepackage[utf8]{inputenc} % allow utf-8 input
\usepackage[T1]{fontenc}    % use 8-bit T1 fonts
\usepackage{hyperref}       % hyperlinks
\usepackage{url}            % simple URL typesetting
\usepackage{booktabs}       % professional-quality tables
\usepackage{amsfonts}       % blackboard math symbols
\usepackage{nicefrac}       % compact symbols for 1/2, etc.
\usepackage{microtype}      % microtypography
\usepackage{xcolor}         % colors
\usepackage{graphicx}
\usepackage{subcaption}
\usepackage{wrapfig}

\title{Pseudo-domains in imaging data improve prediction of future disease status in multi-center studies}

\author{%
      Matthias Perkonigg \\
      Computational Imaging Research Lab (CIR) \\
  Department of Biomedical Imaging and Image-guided Therapy \\
  Medical University of Vienna, Austria\\
  \texttt{matthias.perkonigg@meduniwien.ac.at} \\
   \And
    Peter Mesenbrink\\
      Novartis Pharmaceuticals Corporation \\
      East Hanover, NJ, USA
    \And
    \And
    Alexander Goehler\\
      Novartis Pharmaceuticals Corporation \\
      East Hanover, NJ, USA
    \And
    \And
    Miljen Martic\\
      Novartis Pharma AG \\
      Basel, Switzerland
      \And
    Ahmed Ba-Ssalamah\\
  Department of Biomedical Imaging and Image-guided Therapy \\
  Medical University of Vienna, Austria\\
    \And
    Georg Langs\\
      Computational Imaging Research Lab (CIR) \\
  Department of Biomedical Imaging and Image-guided Therapy \\
  Medical University of Vienna, Austria\\
  \texttt{georg.langs@meduniwien.ac.at} \\
}

\begin{document}

\maketitle

\begin{abstract}
  In multi-center randomized clinical trials imaging data can be diverse due to acquisition technology or scanning protocols. Models predicting future outcome of patients are impaired by this data heterogeneity. Here, we propose a prediction method that can cope with a high number of different scanning sites and a low number of samples per site. We cluster sites into pseudo-domains based on visual appearance of scans, and train pseudo-domain specific models. Results show that they improve the prediction accuracy for steatosis after 48 weeks from imaging data acquired at an initial visit and 12-weeks follow-up in liver disease. 
\end{abstract}

\section{Introduction}
For multi-center studies or randomized clinical trials (RCT) it is important to include a large number of centers to represent the variability of real-world data and to assess if results are generalizable across clinical sites. However, this comes with the challenge of site-specific variability in measurements due to technology such as imaging equipment. Dealing with variability due to different scanner vendors and sites is a common challenge in medical imaging tackled with various approaches such as domain generalization \cite{gulrajani2020search}, unsupervised domain adaptation \cite{Guan2021} or continual learning \cite{Hofmanninger2020a}.

Building site-specific models is often infeasible due to the high number of included sites and low number of samples per site. Here, we propose a method that clusters images across multiple sites to visually homogeneous \textit{pseudo-domains}. Pseudo-domain specific models then accurately account for variability between those clusters, while at the same time exploiting similarity across sites for joint model training. Here, we use pseudo-domain specific models (PDSM) to extract imaging features from longitudinal data. Based on these features a model predicts future outcome of patients from early imaging data.

\section{Data}
Data comprised 74 patients enrolled in a clinical trial for treatment of non-alcoholic steatohepatitis (NASH) \cite{Lucas2020TropifexorResults}. Each patient underwent magnetic resonance imaging (MRI) at an initial visit and two follow-up visits after 12 and 48 weeks, respectively. During each visit, a multi-echo MRI sequence with six echos was acquired and used as imaging data for this study. For the initial visit and 48-week follow-up a liver biopsy was performed and qSteatosis values were calculated. qSteatosis is a score to standardize the histological assessment of steatosis within the liver \cite{Liu2020}. Data was collected from 28 clinical centers, with diverse MRI scanner vendors (GE, Philips and Siemens), scanner models and magnetic field strength (3.0T and 1.5T). Data of the 74 patients is split into training ($n=53$) and test ($n=21$) set.
In the following, the whole data set is refereed as $\mathcal{D} = \{\langle x_1^{(0)}, x_1^{(12)}, x_1^{(48)}, s_1^{(0)}, s_1^{(48)}\rangle, \dots, \langle x_m^{(0)}, x_m^{(12)}, x_m^{(48)}, s_m^{(0)}, s_m^{(48)}\rangle\}$, where $x^{(0)}$, $x^{(12)}$, $x^{(48)}$ are images at baseline and a follow-up after 12 and 48 weeks respectively. $s^{(0)}$ and $s^{(48)}$ are the qSteatosis values at baseline and after 48 weeks of treatment. Two training data sets, extracted from $D$ will be differentiated: $\mathcal{D}_f$ consisted of image-qStetaosis pairs $\langle x, s \rangle$ and was used to train the feature extractors. $\mathcal{D}_p$ consisting of triplets $\langle x^{(0)}, x^{(12)}, s^{(48)}\rangle$ was used to train the prediction model.

\section{Method}
\begin{figure}[t]
  \includegraphics[width=\textwidth]{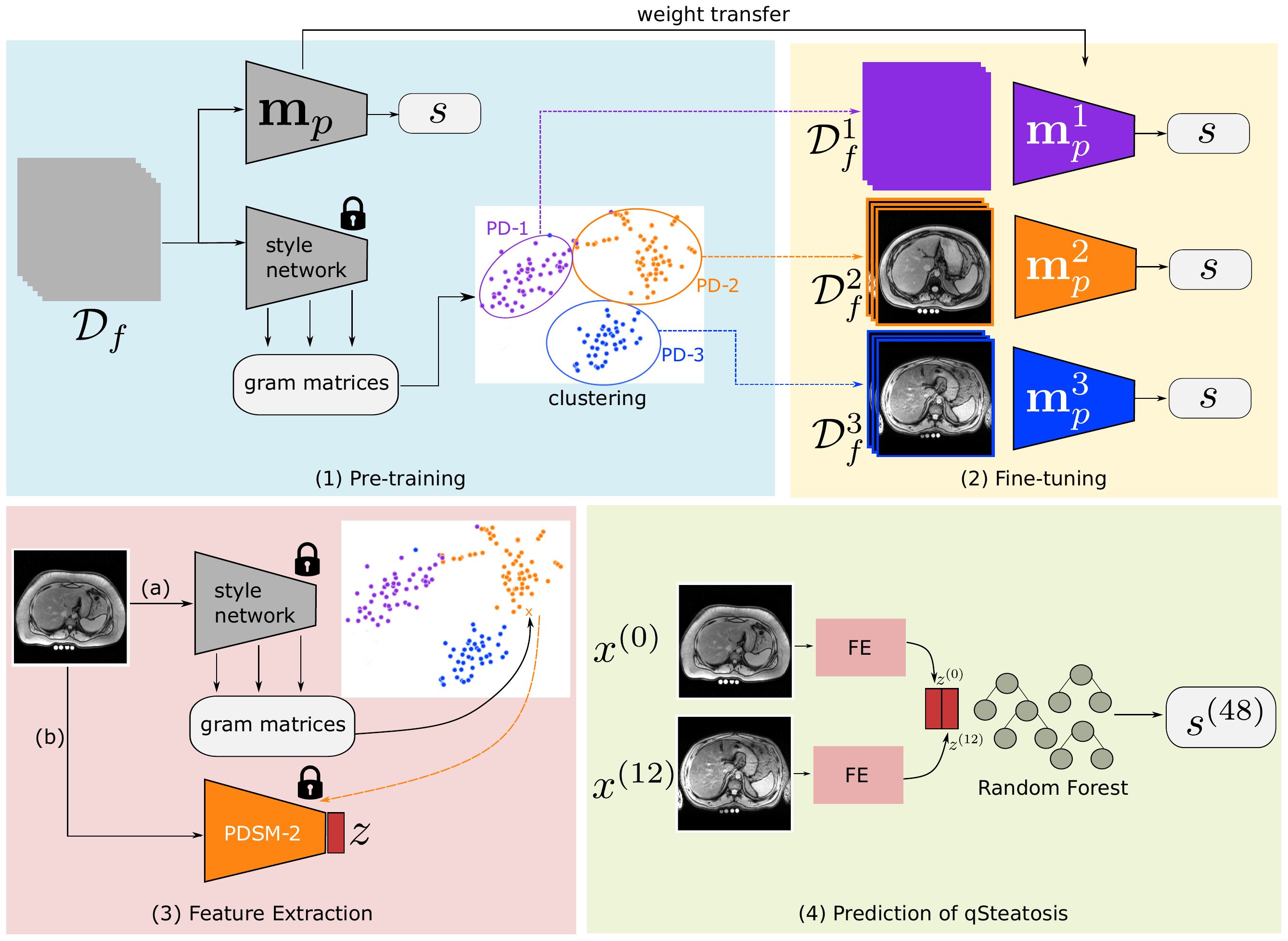}
    \caption{Overview of methodology for predicting future qSteatosis with PDSM. (1) The task network is pre-trained and pseudo-domains are created by clustering in the style embedding space. (2) For each pseudo-domain a PDSM is fine-tuned. (3) Features are extracted from the PDSMs and principle component analysis is applied to reduce the feature dimensionality. (4) A random forest regressor is trained on features from the initial visit and 12-week follow-up to predict qSteatosis at 48-week follow-up.}
    \label{fig:method}
\end{figure}

The proposed method predicts the outcome $s^{(48)}$ from $\langle x^{(0)},x^{(12)} \rangle$. We build pseudo-domain specific models to extract imaging features and afterwards perform the prediction on those features. Figure \ref{fig:method} provides an overview of the methodology. It comprises four steps: 
\paragraph{(1) Pre-training and style embedding} The task network is pre-trained on $\mathcal{D}_f$ to regress qSteatosis from an image $\mathbf{m}_p:x \mapsto s$.
To create a \textit{style embedding}, a style model extracts gram matrices from all images $x \in \mathcal{D}_f$ for a set of pre-defined convolutional layers $\mathcal{L}$. The style embedding $\mathbf{e}(x)$ captures the visual \textit{style} of $x$ \cite{Gatys2016, Hofmanninger2020a}. The style model is pre-trained on a different data set and its weights are fixed during training of the PDSM. Next, k-Means clustering \cite{Lloyd1982} is applied on the style embeddings resulting in a set of $k$ pseudo-domains with distinct visual appearance. For each pseudo-domain a specific training set $\langle\mathcal{D}_f^1, \dots, \mathcal{D}_f^k \rangle$ is sampled from $\mathcal{D}_f$ based on the cluster membership of images.
\paragraph{(2) Fine-tuning of pseudo-domain specific task networks} For each pseudo-domain $d \in \{1\dots k\}$ a separate pseudo-domain specific model $\mathbf{m}_p^d$ is trained by fine-tuning $\mathbf{m}_p$ on the pseudo-domain specific data set $\mathcal{D}_f^d$. 
\paragraph{(3) Feature extraction} For feature extraction an image $x$ is first assigned to one of the pseudo-domains by calculating $\mathbf{e}(x)$ and assigning a pseudo-domain $d$ using the trained k-means clustering. Afterwards, the corresponding model $\mathbf{m}_p^d$ is applied to $x$ and features from the last layer before the classification layer are extracted. Features are extracted for $\forall x \in \mathcal{D}_p$ and principle component analysis (PCA) is applied to reduce the feature dimensions from 512 to 32 to derive the final feature representation $z$.
\paragraph{(4) Outcome prediction} For predicting qSteatosis of a future time point $s^{(48)}$ from $\langle x^{(0)},x^{(12)}\rangle$, features are extracted for baseline $z^{(0)}$ and 12-week follow-up $z^{(12)}$ imaging. $\langle z^{(0)},z^{(12)}\rangle$ are then used to train a random forest regressor $\mathbf{m}_f$ \cite{Breiman2001RandomForests} to predict the qSteatosis value at 48-week follow-up.

\section{Results}

\begin{wrapfigure}{r}{0.5\textwidth}
  \includegraphics[width=0.5\textwidth]{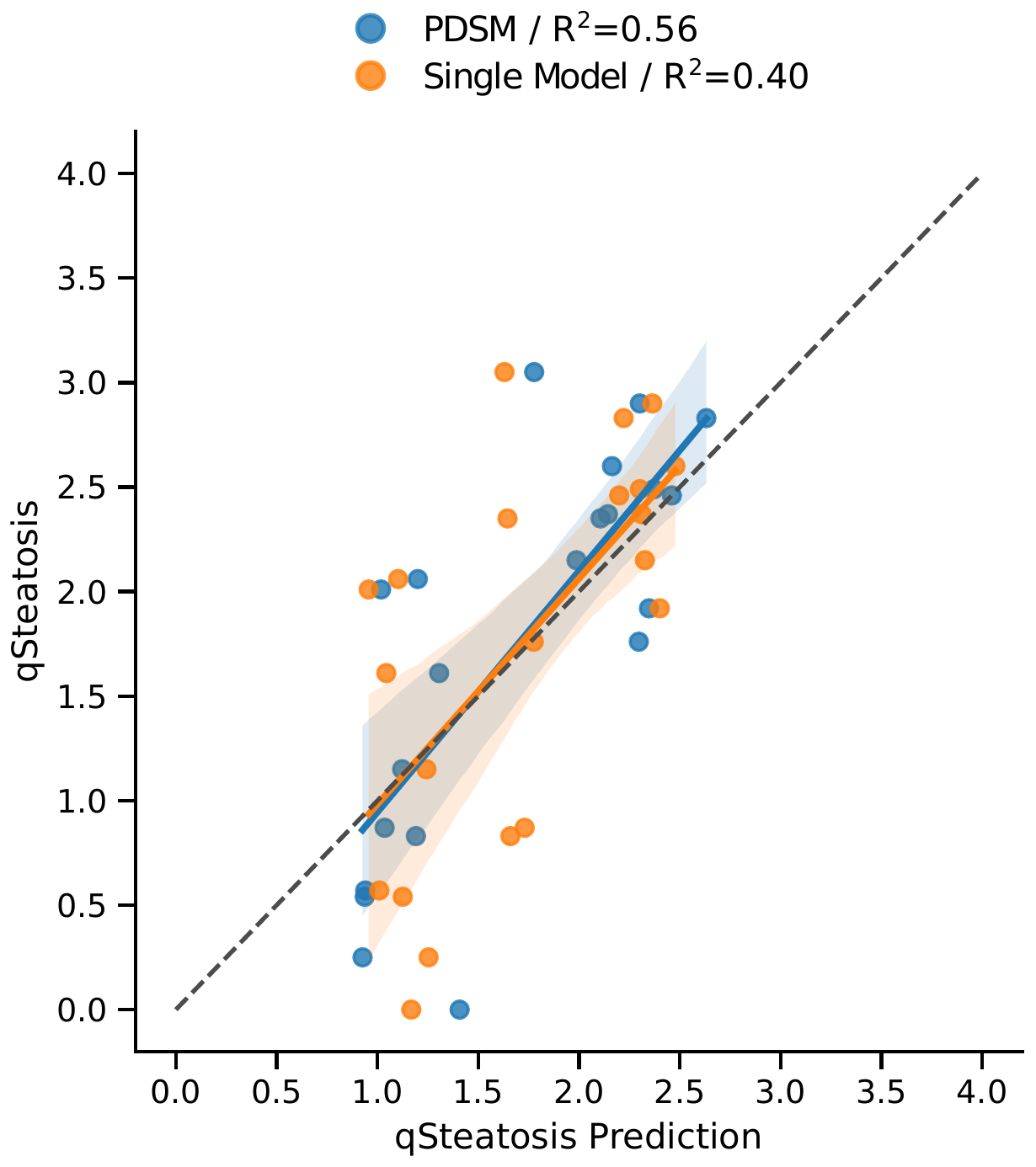}
    \caption{Prediction of qSteatosis improves from $R^2$ of 0.40 to 0.56 when using PDSM.}
    \label{fig:results}
\end{wrapfigure}

The PDSM was trained on 53 patients in the training set with five distinct pseudo-domain clusters resulting in five PDSM models. On the test set of 21 patients we measured $R^2$ and mean squared error (MSE) between the predicted qSteatosis value and the true qSteatosis value at the 48-week follow-up. Using pseudo-domain specific models raised $R^2$ from $0.40$ to $0.56$ compared with using a single model only (see Figure \ref{fig:results}).
MSE was $0.49$ for a single model and $0.36$ for PDSM, showing an advantage of using PDSMs as feature extractors compared to a single model.
To assess whether the dynamics between baseline and 12-week follow up contributed predictive information, we trained a random forest regressor using only the PDSM features extracted from the 12-week follow-up $z^{(12)}$ to predict 48-week outcome. This results in $R^2=0.53$ and an MSE of $0.39$. Results were slightly worse than using both imaging visits ($z^{(0)}$ and $z^{(12)}$) as predictors.

\paragraph{Conclusion} This work shows that pseudo-domains can improve prediction models in multi-center studies with large numbers of sites. Results furthermore demonstrate that the prediction of future outcome from initial imaging data is feasible.  Possible future work will focus on combining multiple PDSMs at inference for previously unseen sites, as they might not match one of the pseudo-domains observed in training.

\section*{Potential negative societal impact}
Models that predict the outcome of a future time-point might lead to the termination of a treatment, if the prediction model is incorrect, potentially worsening an individual patients outcome.

\section*{Acknowledgements}
This study was partially supported by the Novartis Pharmaceuticals Corporation. Furthermore, this study has received partial funding from the European Commission (TRABIT 765148), the Austrian Science Fund (FWF) (P 35189, P 34198), Vienna Science and Technology Fund (WWTF) Project Nr. LS20-065 (PREDICTOME). Part of the computations for research was performed on GPUs donated by NVIDIA.

\small
\bibliographystyle{plain}
\bibliography{neurips_2021}

\end{document}